\documentclass[twocolumn,superscriptaddress,floatfix]{revtex4-2}
\usepackage{graphicx}
\usepackage{latexsym}
\usepackage{amsmath}
\usepackage{amsfonts}
\usepackage{amssymb}
\usepackage{bm}
\usepackage{txfonts}
\newcommand{\fig}[2]{\includegraphics[width=#1]{#2}}
\newcommand{\llangle}{{\langle\!\langle}}
\newcommand{\rrangle}{{\rangle\!\rangle}}
\def\avs{{$A$V$_3$Sb$_5$}}

\def\bM{{\textbf{M}}}

\def\bq{{\textbf{q}}}
\def\bQ{{\textbf{Q}}}
\def\bG{{\textbf{G}}}
\def\ba{{{\bm a}}}

\begin{document}
\title{Interplay of Charge Density Wave and Magnetism on the Kagom\'e Lattice}
	
\author{Yu-Han Lin}
\affiliation{Institute of Theoretical Physics, Chinese Academy of Sciences, Beijing 100190, China}
\affiliation{School of Physical Sciences, University of Chinese Academy of Sciences, Beijing 100049, China}

\author{Jin-Wei Dong}
\affiliation{Anhui Province Key Laboratory of Condensed Matter Physics at Extreme Conditions, High Magnetic Field Laboratory, Chinese Academy of Sciences, Hefei 230031, China}

\author{Ruiqing Fu}
\affiliation{Institute of Theoretical Physics, Chinese Academy of Sciences, Beijing 100190, China}
\affiliation{School of Physical Sciences, University of Chinese Academy of Sciences, Beijing 100049, China}

\author{Xianxin Wu}
\affiliation{Institute of Theoretical Physics, Chinese Academy of Sciences, Beijing 100190, China}

\author{Ziqiang Wang}
\thanks{Corresponding author: wangzi@bc.edu}
\affiliation{Department of Physics, Boston College, Chestnut Hill, MA 02467, USA}
	
\author{Sen Zhou}
\thanks{Corresponding author: zhousen@itp.ac.cn}
\affiliation{Institute of Theoretical Physics, Chinese Academy of Sciences, Beijing 100190, China}
\affiliation{School of Physical Sciences, University of Chinese Academy of Sciences, Beijing 100049, China}

\begin{abstract}
Motivated by the recent discovery of charge density wave (CDW) order in the magnetic kagom\'e metal FeGe, we study the single-orbital $t$-$U$-$V_1$-$V_2$ model on the kagom\'e lattice, where $U$, $V_1$, and $V_2$ are the onsite, nearest neighbor, and next-nearest-neighbor Coulomb interactions, respectively.
When the Fermi level lies in the flat band, the instability toward ferromagnetic (FM) order gives rise to a FM half-metal at sufficiently large onsite $U$. 
Intriguingly, at band filling $n=17/24$, the Fermi level crosses the van Hove singularity of the spin-minority bands of the half-metal.
We show that, due to the unique geometry and sublattice interference on the kagom\'e lattice at van Hove singularity, the inter-site Coulomb interactions $V_1$ and $V_2$ drive a real and an imaginary bond-ordered $2a_0 \times 2a_0$ CDW instability, respectively.
The FM loop current CDW with complex bond orders is a spin-polarized Chern insulator exhibiting the quantum anomalous Hall effect.
The bond fluctuations are found to be substantially enhanced compared to the corresponding nonmagnetic kagom\'e metals at van Hove filling, providing a concrete model realization of the bond-ordered CDWs, including the FM loop current CDW, over the onsite charge density ordered states.
When the spins are partially polarized at an intermediate $U$, we find that the interplay of CDW and magnetism enables the formation of real and complex bond-ordered CDWs, and the CDW transition is accompanied by a substantial enhancement in the ordered magnetic moments.
These findings provide physical insights for the emergence of $2a_0 \times 2a_0$ CDWs and their interplay with magnetism on the kagom\'e lattice, and capture the essential physics observed experimentally in FeGe.
\end{abstract}
\maketitle

\textit{Introduction.} - Kagom\'e metals \avs\ ($A$ = K, Rb, Cs) \cite{Ortiz-PRM19} have attracted enormous attention recently as a fertile playground for exploring novel states of matter \cite{Neupert-np22, Jiang-nsr22, Wilson-nrm24} due to the interplay between lattice geometry, band topology, and electron correlation inherent on the kagom\'e lattice.
A rich array of correlated and topological electronic states has been observed \cite{Ortiz-prl20, Ortiz-prm21, HXLi-prx21, Jiang-nm21, Zhao-nat21, Ortiz-prx21, Chen-prx21, Ortiz-prb21, Shumiya-prb21, Hong-np22, ChenHui-nat21, HCLei-cpl21, LPNie-nat2022, ZCJiang-NanoLett23, Xing-nat24}, including a novel charge density wave (CDW) with $2a_0 \times 2a_0$ in-plane periodicity below $T_\text{cdw}$, rotation symmetry breaking, charge-stripe order, electronic nematicity, and superconductivity exhibiting pair density wave order.
Central to the physics of \avs\ is the evidence for spontaneous time-reversal symmetry (TRS) breaking in the exotic CDW state \cite{Jiang-nm21, Xing-nat24, Mielke-Nat22, WuQiong-PRB22, XuYishuai-NatPhys22, GuoChunyu-Nat22, YangShuoYing-SciAdv20, ChenXH-PRB21}, despite the absence of local moment \cite{Kenney-IOP21} and itinerant magnetism \cite{Ortiz-PRM19, Ortiz-prm21}. 
A promising candidate for the TRS breaking is the bond CDW with persistent loop current (LC) order \cite{LinYuPing-PRB21, FengXilin-SciBull21, Denner-PRL21, SZ-NC22}, a long-sought-after quantum order also relevant for the quantum anomalous Hall (QAH) insulator \cite{Haldane-PRL88} and the pseudogap phase of cuprate superconductors \cite{Affleck-PRB1988, Varma-PRB1997, Nayak-PRB00, WenXG-RevMod06}. 
The physical origin, model realization, and broader implications of the proposed LC order have been a subject under intensive research \cite{LinYuPing-PRB21, YanBinghai-PRL21, SZ-NC22,Dong-PRB23, Li-PRL24, Christensen-PRB21, Park-PRB21, FengXilin-PRB21}.

More recently, a CDW order with identical $2a_0\times 2a_0$ in-plane periodicity was discovered in the magnetic kagom\'e metal FeGe below $T_\text{cdw}$$\sim$110 K \cite{PCDai-Nature22, JXYin-PRL22, MYi-NatPhys23}, deep inside the A-type antiferromagnetic (AFM) phase ($T_N$$\sim$410 K).
Here the magnetic moments order ferromagnetically (FM) within each kagom\'e layer with the out-of-plane moments anti-aligned between adjacent layers \cite{Watanabe-JPSJ66, Beckman-PS72}. 
Moreover, the ordered magnetic moments increases substantially at the onset of the CDW transition \cite{PCDai-Nature22}, indicating strong coupling between CDW and magnetism. 
Though recent theoretical calculations \cite{arXiv22-Setty, ZhouHJ-PRB23, XGWan-CPL23, YWang-PRM23, Chang-acs23, JPLiu-CPL24, arXiv23-HXiang} and experimental investigations \cite{MYi-NatPhys23, arXiv23-FengDL, arXiv23b-FengDL, Lee-NC23, arXiv24-XQiu, arXiv24-XYin, ChenXH-PRR24, WangAF-PRL24, Wenzel-PRL24, DaiPC-PRL24} suggest a CDW mechanism in FeGe different than \avs, the presence of van Hove (vH) singularities in the vicinity of the Fermi level revealed in angle-resolved photoemission spectroscopy (ARPES) \cite{MYi-NatPhys23}, together with the observed anomalous Hall effect \cite{PCDai-Nature22} and pronounced edge modes \cite{JXYin-PRL22}, hint at possible connections between the CDWs in magnetic FeGe and nonmagnetic \avs\  kagom\'e metals.
Here, we unite the physics of flat band FM and vH singularity CDW, investigate their interplay on the kagom\'e lattice, and uncover the possible implications to FeGe.

\begin{figure*}
\begin{center}
\fig{7.in}{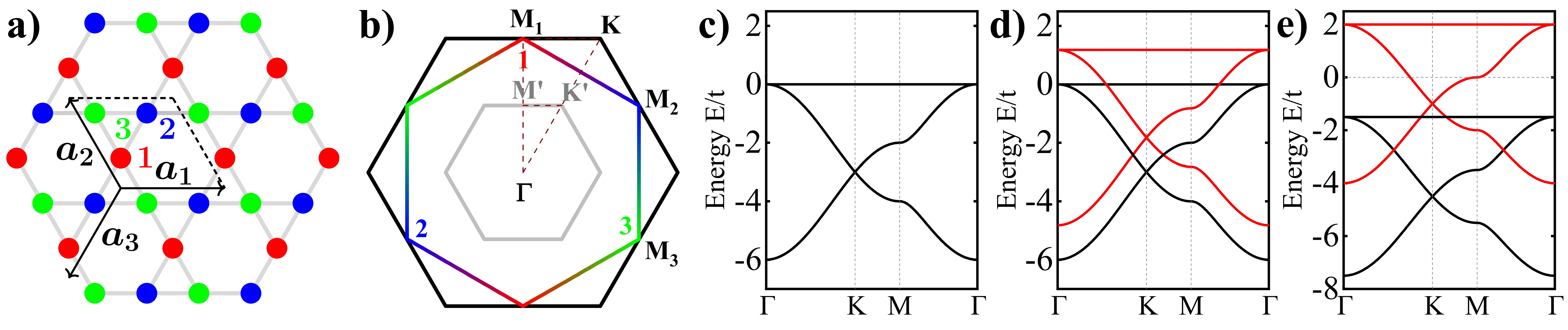}
\caption{(a) Kagom\'e lattice with three sublattices denoted by red (1), blue (2), and green (3) circles.
The two basis lattice vector are $\ba_1 = (1,0)$ and $\ba_2= (-{1\over 2}, {\sqrt{3} \over 2})$, with the third direction follows $\ba_3$ = $-\ba_1$ $-\ba_2$.
The corresponding reciprocal lattice vectors are $\bG_1 = (0, {4\pi \over \sqrt{3}})$, $\bG_2 = (-2\pi, -{2\pi \over \sqrt{3}})$, and $\bG_3$ = $-\bG_1$ $-\bG_2$.
(b) FS at vH filling with sublattice contents displayed by color of lines and numbers.
The three independent vH points are labeled by $\bM_{1,2,3}={1\over 2} \bG_{1,2,3}$, respectively.
Black and gray lines enclose the original and $2a_0 \times 2a_0$ reduced BZ.
(c) Tight-binding dispersion at filling $n=17/24$ in the original BZ, with the Fermi level passing through the flat band.
Electronic structures along the high-symmetry path in the original BZ for (d) the partially polarized FM at $U=3.6$ and (e) the fully polarized FM at $U=6$.
Black and red curves denote, respectively, the spin-majority ($\uparrow$) and spin-minority ($\downarrow$) bands.} \label{fig1}
\end{center}
\vskip-0.5cm
\end{figure*}

To this end, we consider the minimal single-orbital model on the kagom\'e lattice depicted in Fig. \ref{fig1}a.
The Fermi surface (FS) at vH filling $n=5/12$ is the hexagon in Fig. \ref{fig1}b that connects the three independent vH points labeled by $\bM_{1,2,3}$ at the Brillouin zone (BZ) boundary.
These vH points with divergent density of states are perfectly nested by the $2a_0 \times 2a_0$ wave vectors $\bQ_\alpha ={1\over 2}\bG_\alpha$, where $\alpha=1,2,3$ and $\bG_\alpha$ is the reciprocal wave vector.
Since the Bloch state at vH point $\bM_\alpha$ is exclusively localized on the $\alpha$th sublattice due to the sublattice interference \cite{LiJX-prb12, Thomale-prb12, Thomale-prl13, WangQH-prb13, SZ-NC22}, as shown in the sublattice-resolved FS in Fig. \ref{fig1}b, the leading $2a_0 \times 2a_0$ instability is believed to be associated with a bond order that connects different sublattices.
Therefore, while onsite Coulomb interaction $U$ acting on the same sublattice is obstructed, inter-site Coulomb interactions are important for the formation of $2a_0 \times 2a_0$ CDWs. 
Weak-coupling and mean-field theories find that inter-site Coulomb interactions $V_1$ on nearest neighbor (nn) bonds and $V_2$ on next-nn (nnn) bonds indeed drive real and complex bond-ordered CDW with LC order \cite{Dong-PRB23}.
This picture captures the most essential physics of CDW in nonmagnetic \avs.

We propose that for the magnetic FeGe, the Fermi level in the high-temperature symmetric phase lies instead inside the flat band of the Fe $d$ electrons (Fig. \ref{fig1}c), which is conjectured based on density-functional theory (DFT) calculations and ARPES observations \cite{MYi-NatPhys23}. 
The onsite Hubbard $U$, assisted by the divergent uniform spin susceptibility tied to the flat band, drives a FM ordered state with spin-split bands shown in Fig. \ref{fig1}d. 
Intriguingly, at band filling $n=17/24$ and for a sufficiently large $U$, the strong FM ordered phase becomes a half-metal (Fig. \ref{fig1}e) where the spin-majority ($\uparrow$) bands are pushed completely below the Fermi level that passes through the vH singularity of the spin-minority ($\downarrow$) bands, \textit{i.e.} $n_\uparrow=1$ and $n_\downarrow=5/12$. 

The spin-polarized bond fluctuations turn out to be substantially enhanced compared to the spin-degenerate nonmagnetic kagom\'e metals at vH filling \cite{arXiv24-FuRQ}. 
The leading instability of the FM half-metal is toward nn real and nnn imaginary bond orders rather than the onsite charge density ordered (CDO) states \cite{arXiv24-FuRQ, arXiv24-ZhanJ}. 
When all particle-hole channels in the Wick contraction are treated on equal footing, we show that the inter-site Coulomb interactions drive the FM ordered state into spin-polarized $2a_0 \times 2a_0$ real and complex CDWs with LC order, providing a  concrete model realization of bond-ordered CDW and LC order for physical spin-1/2 electrons.

For intermediate $U$, which is expected for FeGe, we show that partially polarized FM order emerges with vH points of spin-minority bands far below the Fermi level.
Remarkably, we discover a mechanism rooted in the interplay between CDW and magnetism that drives the vH points to the Fermi level and enables bond-ordered CDWs induced by inter-site Coulomb interactions. 
The ordered magnetic moments can be substantially enhanced at the onse of CDW transition, providing a plausible explanation for the experimental findings in FeGe \cite{PCDai-Nature22}.

\textit{Model and mean-field theory.} - We consider the effective single-orbital $t$-$U$-$V_1$-$V_2$ model on the kagom\'e lattice
\begin{equation}
H =  -t \hspace{-0.07cm} \sum_{\langle i,j\rangle, \sigma}  c^\dagger_{i\sigma} c_{j\sigma}
+U \hspace{-0.07cm} \sum_i \hat{n}_{i\uparrow} \hat{n}_{i\downarrow}
+V_1 \hspace{-0.07cm} \sum_{\langle i,j\rangle} \hat{n}_i \hat{n}_j
+V_2 \hspace{-0.07cm} \sum_{\langle\langle i,j \rangle\rangle} \hat{n}_i \hat{n}_j,
\label{tUV}
\end{equation}
where $c^\dagger_{i\sigma}$ creates a spin-$\sigma$ electron on site $i$, the electron density operators $\hat{n}_{i\sigma} = c^\dagger_{i\sigma} c_{i\sigma}$ and $\hat{n}_i = \sum_\sigma \hat{n}_{i\sigma}$.
The strengths of Coulomb interactions in Eq. (\ref{tUV}) should be considered as phenomenological input parameters, though their orbital averaged values in a five-$d$-orbital model for FeGe are recently estimated to be around $U$ = 4.12 eV and $V$ = $(V_1$, $V_2)$ = (1.41, 1.22) eV \cite{Bernevig-prb25}.
Hereinafter, we set nn hopping $t\equiv 1$ as the energy unit.
Decoupling the onsite Coulomb interaction in terms of local magnetic moment $\hat{m}^\eta_i$ = $\sum_{\sigma \sigma'} c^\dagger_{i\sigma} \tau^\eta_{\sigma \sigma'} c_{i\sigma'}$ ($\eta=x,y,z$), and the inter-site Coulomb interactions in terms of bonds $\hat{\chiup}^\mu_{ij}$ = $\sum_{\sigma \sigma'} c^\dagger_{i\sigma} \tau^\mu_{\sigma \sigma'} c_{j\sigma'}$ ($\mu=0,x,y,z$), with $\tau^{\eta/\mu}$ the corresponding Pauli matrix, one obtains the mean-field Hamiltonian
\begin{align}
H_\text{MF}=&-t\sum_{\langle i,j\rangle} \hat{\chiup}^0_{ij} -{U\over 4} \sum_{i,\eta} \left[ 2m^\eta_i \hat{m}^\eta_i - \left(m^\eta_i \right)^2 \right] \nonumber \\
&- {V_1\over 2} \sum_{\langle i,j\rangle, \mu} \left[ \left(\chiup^\mu_{ij} \right)^* \hat{\chiup}^\mu_{ij} +h.c. - \left|\chiup^\mu_{ij} \right|^2 \right] \nonumber \\
&- {V_2\over 2} \sum_{\llangle i,j\rrangle, \mu} \left[ \left(\chiup^\mu_{ij} \right)^* \hat{\chiup}^\mu_{ij} +h.c. - \left|\chiup^\mu_{ij} \right|^2 \right]. \label{HMF}
\end{align}
The order parameters, magnetic moment $m^\eta_i =\langle \hat{m}^\eta_i \rangle$ and bonds $\chiup^\mu_{ij} =\langle \hat{\chiup}^\mu_{ij} \rangle$, are to be determined self-consistently by minimizing the state energy.

\begin{figure*}
\begin{center}
\fig{7.in}{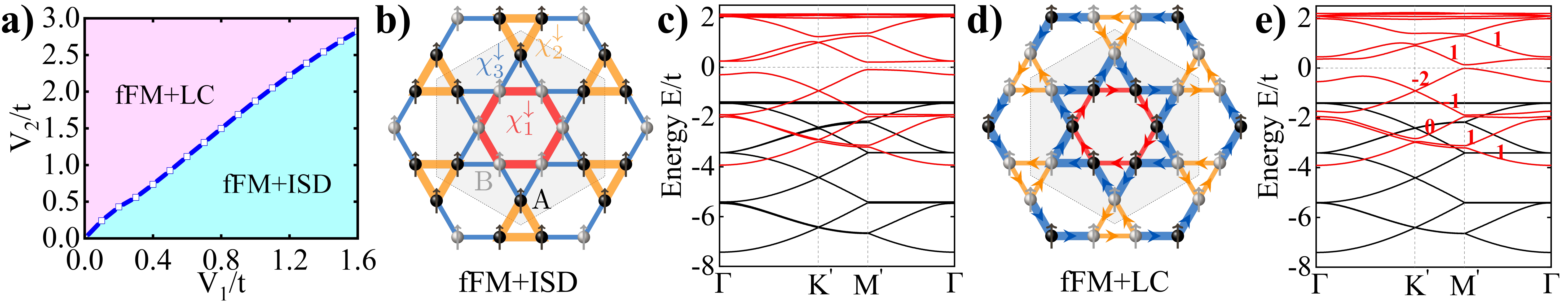}
\caption{(a) Ground state phase diagram of the kagom\'e lattice $t$-$U$-$V_1$-$V_2$ model at filling $n=17/24$ with fixed onsite $U=6$.
Solid line represents the phase boundary of first-order transition between $f$FM+ISD and $f$FM+LC.
(b) Schematics picture and (c) band dispersion in the reduced BZ for the $f$FM+ISD at $V=(1.5, 1)$.
(d) Schematics picture and (e) band dispersion for the $f$FM+LC at $V=(1.2, 2.8)$.
In (b) and (d), the shaded areas enclosed by dotted lines denote the enlarged $2a_0 \times 2a_0$ unit cell, the width of bonds indicate the hopping of spin-minority electrons on nn bonds, the bond arrows display the flowing direction of LCs, and the site arrows show the FM moments.
Numbers in (e) denote the Chern numbers of corresponding spin-minority bands.} \label{fig2}
\end{center}
\vskip-0.5cm
\end{figure*}

We note that in the mean-field theory used here, the direct Hartree terms of the interactions depending on the electron density $n_i =\langle \hat{n}_i \rangle$ are neglected to avoid double-counting, since their contributions are already considered in DFT \cite{Dong-PRB23, JiangK-PRB16}.
These terms are expected to be insignificant in these kaogm\'e metals, since no large charge disproportionation has been observed.
Moreover, the symmetry preserving bond corrections, \textit{i.e.}, the real uniform components of the bond order parameters $\chiup^\mu_{ij}$, are subtracted in the interactions, such that the vH singularities and the band structure are maintained without normalizing the bare band parameters \cite{Dong-PRB23, ZhangY-PRB20}.
This is consistent with the fact that vH singularities at the $\bM$ points have been established by ARPES in both \avs\ and FeGe.
Implementing this subtraction scheme, the correlation effects in the self-consistent mean-field theory will only generate spontaneous symmetry-breaking states.

We stay at band filling $n=17/24$, at which the Fermi level of free electrons passes through the flat band, as shown in the tight-binding band dispersion in Fig. \ref{fig1}c.
The presence of flat band at Fermi level gives rise to divergent bare susceptibilities at wave vector $\bq=0$, and consequently leads to collinear FM at any infinitesimal $U$.
Though spins in the studied model have global SU(2) symmetry, we restrict them to order along the $c$-axis, in connection to FeGe.
Consequently, bond orders $\chiup_{x,ij}$ and $\chiup_{y,ij}$ vanish on all bonds.
It is thus more convenient to define the spin-dependent bonds, $\chiup^\sigma_{ij} = {1\over 2}(\chiup^0_{ij} +\sigma \chiup^z_{ij})$, which we use hereinafter.
Furthermore, we consider quantum states described by the Hamiltonian in Eq. (\ref{HMF}) to be periodic with an enlarged $2a_0 \times 2a_0$ unit cell, which allows investigations on the interplay between $\bq=0$ magnetism and $2a_0 \times 2a_0$ CDWs.
In addition, we focus on states invariant under $C_6$ rotation and satisfying the current continuity condition \cite{Dong-PRB23}, which implies that there are, independently, two lattice sites, three nn bonds, and three nnn bonds within each $2a_0 \times 2a_0$ enlarged unit cell.
In order to obtain all the possible states, we use different initial conditions for solving the self-consistent equations numerically.
When multiple converged states exist for a given set of interactions, we compare the state energies to determine the true mean-field ground state.

\textit{FM LC CDWs} - 
The fully polarized FM ($f$FM) phase at sufficiently large $U$ is a half-metal, with the spin-majority ($\uparrow$) bands pushed completely below the Fermi level.
Remarkably, the Fermi level exactly passes through the vH points of the spin-minority ($\downarrow$) bands, as shown by the band dispersion at $U=6$ displayed in Fig. \ref{fig1}e.
Therefore, the fully polarized $t$-$U$-$V_1$-$V_2$ model at a fixed large $U$ and filling $n=17/24$ provides effectively a spinless counterpart of the spinful $t$-$V_1$-$V_2$ model at vH filling studied in Ref. \cite{Dong-PRB23}.
Fig. \ref{fig2}a presents the mean-field phase diagram at fixed $U=6$ in the plane spanned by $V_1$ and $V_2$.
Indeed, nn $V_1$ and nnn $V_2$ drives, respectively, real and complex $2a_0 \times 2a_0$ bond-ordered CDW coexisting with $f$FM induced by the large onsite $U$.

The ground state at $V=(1.5, 1)$ is schematized in Fig. \ref{fig2}b.
The electron densities on the two inequivalent sites, $A$ and $B$, are 1.404 and 1.430, and both of them are fully spin polarized.
Consequently, the spin-majority ($\uparrow$) electrons are completely localized on lattice sites, with $\chiup^\uparrow_{ij}=0$ on all bonds.
The nn bonds for spin-minority ($\downarrow$) electrons are real and  $\chiup^\downarrow$ = (0.264, 0.268, 0.172), with the stronger nn bonds form the inverse Star-of-David (ISD) pattern displayed in Fig. \ref{fig2}b.
This coexistent state of $f$FM and real bond-ordered CDW is thus referred to as $f$FM+ISD from now on.
Fig. \ref{fig2}c plots the corresponding band dispersion, where the vH points of spin-down bands near the Fermi level are fully gapped by the $2a_0\times 2a_0$ CDW.

The coexistent state of $f$FM and complex bond-ordered CDW obtained at $V=(1.2, 2.8)$ is illustrated in Fig. \ref{fig2}d, with $n_A$=1.420, $n_B$=1.413, and $\chiup^\downarrow$ = (0.197+0.028$i$, 0.194$-$0.029$i$, 0.241+0.028$i$).
We call this state as $f$FM+LC hereinafter, since the complex bonds of spin-down electrons generate spin-polarized LCs circulating on the kagom\'e lattice.
Consequently, the spin-down bands are topologically nontrivial \cite{Dong-PRB23}.
The band dispersion is presented in Fig. \ref{fig2}e, with the Chern numbers of the isolated spin-down bands indicated by numbers.
Note that while the Chern number of an individual band may change its value as $V_{1,2}$ varies in the $f$FM+LC regime in the phase diagram, the total Chern number of all occupied bands remains unchanged.
As a result, $f$FM+LC is a spin-polarized Chern insulator exhibiting QAH effect.

\begin{figure}
	\begin{center}
		\fig{3.4in}{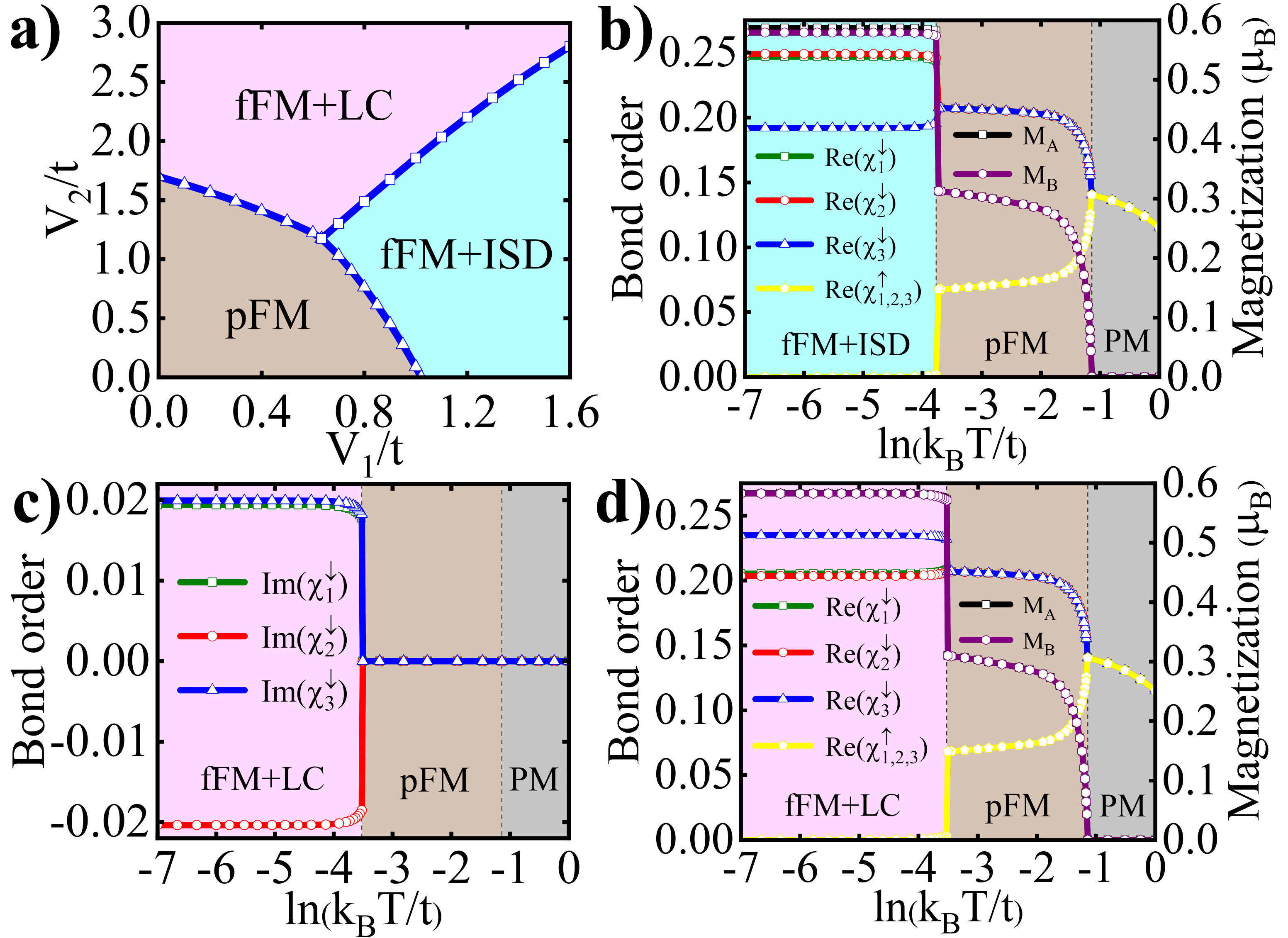}
		\caption{(a) Ground state phase diagram in the $V_1$-$V_2$ plane at fixed $U=3.6$.
			Solid lines represents the phase boundary of first-order transitions.
			Temperature evolution of (b) $f$FM+ISD at $V=(1, 1)$, and (c,d) $f$FM+LC at $V=(0.8, 2)$.
		} \label{fig3}
	\end{center}
	\vskip-0.5cm
\end{figure}

\textit{CDW enhanced magnetism} - Next, we set onsite $U=3.6$ where the magnetic ground state is a partially polarized FM ($p$FM) with ordered moment $m$ = 0.328 $\mu_B$.
The band dispersion is shown in Fig. \ref{fig1}d.
Clearly, the vH points of spin-minority bands are driven close to, but still quite far below the Fermi level, due to the weak FM splitting in $p$FM.
If CDW and magnetism were decoupled, the magnetization would persist unaltered when inter-site Coulomb interactions are introduced, and the latter of any physical magnitude would fail to gap out the vH points far below the Fermi level, thus preventing the emergence of $2a_0 \times 2a_0$ CDWs. 
Instead, due to the interplay between CDW and magnetism, the system would rather pay some energy to fully polarize the spins and drive the vH points close to the Fermi level, such that the vH points can be gapped out to gain more energy in the formation of $2a_0 \times 2a_0$ CDWs.
Indeed, as shown in Fig. \ref{fig3}a at fixed $U=3.6$, while $p$FM remains as the ground state in the regime where inter-site interactions are weak, real and complex $2a_0 \times 2a_0$ CDWs set in when $V_1$ and $V_2$ are sufficiently strong, resulting in $f$FM+ISD and $f$FM+LC in the phase diagram.

The substantial enhancement of ordered magnetic moments at the onset of CDW transition is further illustrated in the temperature evolution of $f$FM+ISD at $V=(1, 1)$ in Fig. \ref{fig3}b and $f$FM+LC at $V=(0.8, 2)$ in Figs. \ref{fig3}c and \ref{fig3}d.
At high temperatures, the system is a paramagnetic (PM) metal at both sets of interactions, with $m=0$ on all sites and $\chi^\uparrow_{ij} =\chi^\downarrow_{ij}$ on all bonds.
As temperature reduces, $p$FM with spins partially polarized sets in via a continuous phase transition.
The spin-up and spin-down bonds start to deviate from each other.
This phase transition breaks the time-reversal symmetry but preserves the translation symmetry.
The latter is broken at a lower temperature, where a first-order phase transition takes place to the coexisting state of $f$FM and $2a_0 \times 2a_0$ bond-ordered CDW, \textit{i.e.}, $f$FM+ISD at $V=(1, 1)$ and $f$FM+LC at $V=(0.8, 2)$.
This transition is accompanied by a substantial enhancement in the ordered magnetic moments \cite{PCDai-Nature22} and the energy separation between spin-majority and spin-minority bands.
The latter may induce a sudden spectral weight transfer detectable in optical spectroscopy \cite{Wenzel-PRL24, arXiv24-XQiu}. 

The interplay of CDW and magnetism near flat band filling $n$ = 17/24 is further elaborated in the Supplementary.
We show that the interplay tends to drive the vH points of the spin-minority bands to the Fermi level to gain more energy in the opening of the CDW gap, in competition with the cost to alter the magnetic state.
The upward (downward) movement of the vH points reduces (increases) the filling of the spin-minority bands, thus results in an enhancement (reduction) in the ordered magnetic moments.
We thus argue FeGe to be an intermediately correlated kagom\'e metal at a flat band filling close to $n$ = 17/24.
It becomes a $p$FM at temperatures between $T_N$ and $T_\text{cdw}$, with the vH points of the spin-minority bands lying underneath the Fermi level.
The interplay of CDW and magnetism leads to a significant enhancement in the ordered magnetic moments at $T_\text{cdw}$, the onset of transition to either real or complex bond-ordered CDW with LC order.

\begin{figure}
	\begin{center}
		\fig{3.4in}{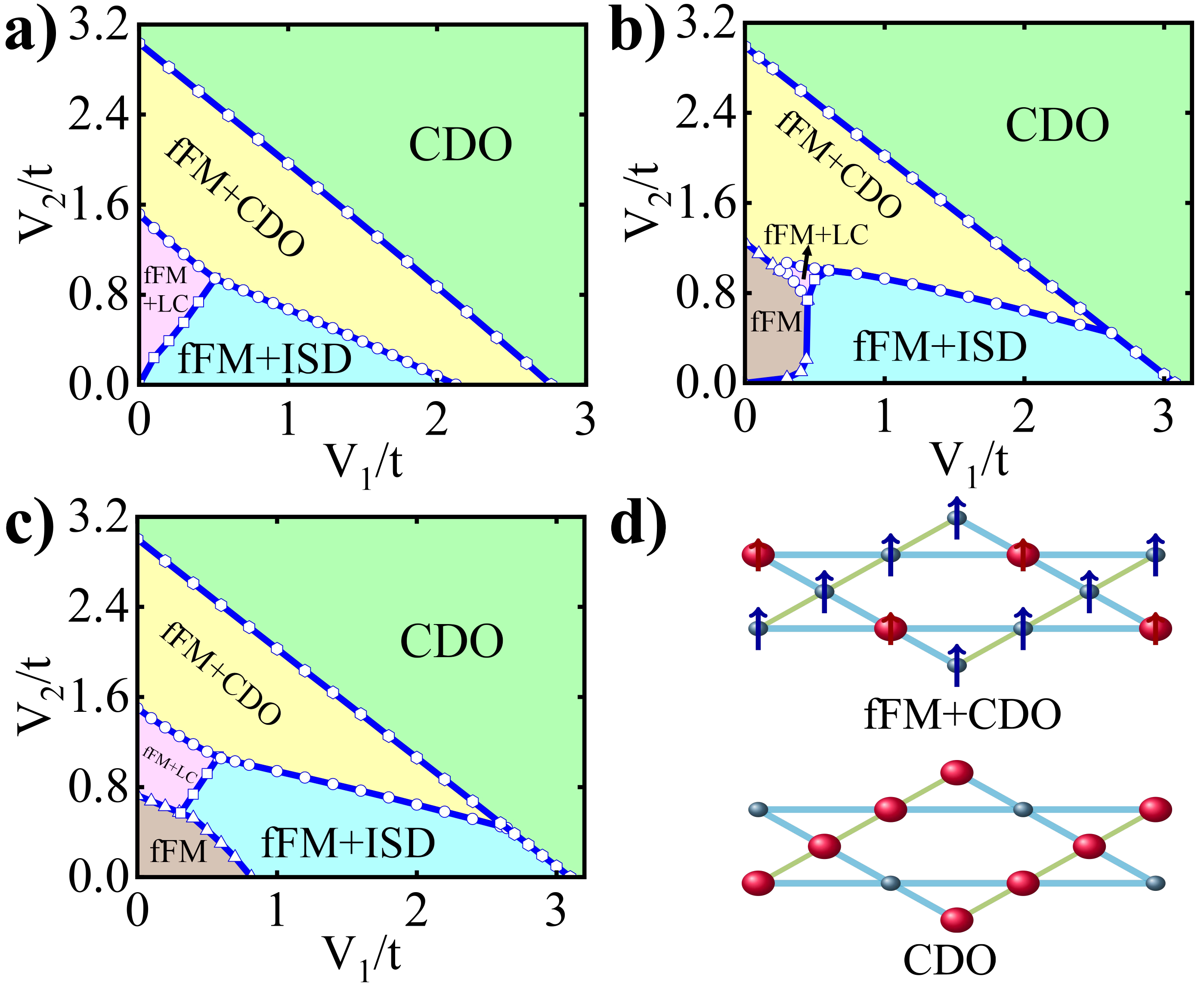}
		\caption{Ground state phase diagrams of the kagom\'e lattice $t$-$U$-$V_1$-$V_2$ model at fixed $U=6t$ after including (a) the Hartree terms and (b) both Hartree terms and symmetry-preserving bond corrections, i.e., the full mean-field calculation.
			(c) Similar to (b), but with a finite hopping $t'=0.015t$ on the nnn bonds.
			(d) Schematic comparison between $f$FM+CDO and CDO states.
			Size of spheres denote the local electron density and the site arrows represent the FM moments.}
		\label{fig4}
	\end{center}
	\vskip-0.5cm
\end{figure}

\textit{Rivalry with onsite CDO.} - So far, the mean-field theory used in this work has subtracted the contributions from onsite Hartree terms and the symmetry preserving bond corrections.
Though these subtractions are reasonable and physical, as we discussed above, it is desirable to investigate theoretically if the exotic LC order can develop spontaneously in the concrete model described in Eq. (\ref{tUV}) within a complete and self-contained mean-field theory.

The perfectly nested FS and its sublattice features shown in Fig. \ref{fig1}b inform us that the vH filled kagom\'e lattice has instabilities at $2a_0 \times 2a_0$ wavevectors $\bq=\bM$ associated with bond orders connecting two different sublattices.
However, the FS also has a $\bq=0$ instability toward onsite charge density order, due to the divergent density of states of the vH singularity.
Therefore, the various Hartree terms resulting from decoupling the Coulomb interactions in terms of the onsite electron densities can drive a unique $\bq=0$, \textit{i.e.}, intra-unit-cell or $1a_0 \times 1a_0$, order of the onsite charge density on different sublattices.
The energy gain of such onsite CDO is dominated by the classical electrostatic potential, and prevails over bond-ordered CDWs when the inter-site interactions are strong.
Furthermore, nnn $V_2$ introduces dynamically a nonzero nnn hopping integral and thus weakens or even removes the vH singularities at the $\bM$ points, further discouraging the emergence of bond-ordered real or complex CDW with LC.
Indeed, including the Hartree terms that were omitted in Ref. \cite{Dong-PRB23} to avoid double-counting beyond the DFT calculations, the LC CDW  fails to be the ground state in the entire phase diagram in the $t$-$V_1$-$V_2$ model at vH filling, as shown in the Supplementary.
This raises the question whether the LC order can emerge spontaneously within the framework of a microscopic model on the kagom\'e lattice.

Intriguingly, we find that the $t$-$U$-$V_1$-$V_2$ model provides a self-contained model realization of LC order in the ground state in the physical parameter regime of the phase diagram.
First, including the Hartree terms, we obtained the ground state phase diagram shown in Fig. \ref{fig4}a at $n=17/24$ and $U=6$.
Compared to Fig. \ref{fig2}a, the regime with strong inter-site interactions is now taken over by onsite CDO or its coexistence with $f$FM.
These states do not break translation symmetry, but break rotation symmetry down to $C_2$ and are therefor electronic nematic states.
The fact that the LC order survives in the phase diagram is in sharp contrast to the previous studies of the $t$-$V_1$-$V_2$ model at vH filling where the LC order is completely removed by the inclusion of the Hartree terms (Supplementary).
The reason for this can be understood by noting that the LC order is induced from the spin-minority bands of the $f$FM, which weakens the electrostatic potential of the Hartree terms.
This finding is consistent with the recent random phase approximation \cite{arXiv24-FuRQ} and functional renormalization group \cite{arXiv24-ZhanJ} studies of the spinless $t$-$V_1$-$V_2$ model at vH filling.

The phase diagram obtained by complete mean-field calculations is displayed in Fig. \ref{fig4}b, where some of the area for $f$FM+LC is further converted to the $f$FM metal.
Nevertheless, LC order survives in a small parameter regime in the $t$-$U$-$V_1$-$V_2$ model even when magnetism, onsite CDO, and bond orders are treated on equal footing.
The regime for LC order can be enlarged via introducing a finite nnn hopping $t'$, as shown in Fig. \ref{fig4}c for $t'=0.015$.
Interestingly, $f$FM+CDO and CDO illustrated in Fig. \ref{fig4}d differ to each other dramatically, as the two equivalent sites in the unit cell have smaller electron density in $f$FM+CDO but larger density in CDO.
This may lead to different dynamical behaviors under ultrafast optical pump pulses \cite{JMoore24}.

\textit{Summary and discussions.} - The effective single-orbital $t$-$U$-$V_1$-$V_2$ model on the kagom\'e lattice is studied within mean-field theories at $n=17/24$, at which the noninteracting Fermi level passes the flat bands.
The system becomes a $f$FM half-metal at sufficiently large onsite $U$, with the spin-minority bands realizing effectively a spinless model at vH filling.
Consequently, ISD and LC with real and complex $2a_0 \times 2a_0$ bond orders are induced by inter-site Coulomb interactions, giving rise to $f$FM+ISD and $f$FM+LC.
The former is a topologically trivial insulator, whereas $f$FM+LC is a spin-polarized Chern insulator exhibiting QAH effect.
We have thus provided a concrete model realization of the emergent $2a_0 \times 2a_0$ bond-ordered CDWs with or without LC order inside the FM ordered state.

We showed that intermediately correlated magnetic kagom\'e metals, such as FeGe, can realize partially polarized $p$FM at temperatures between $T_N$ and $T_\text{cdw}$, with the vH points of the spin-minority bands lying far underneath the Fermi level.
The interplay of CDW and magnetism drives the vH points close to the Fermi level below $T_\text{cdw}$ at reasonable inter-site interactions, thus enables the formation of $2a_0 \times 2a_0$ CDWs.
The ordered magnetic moments are enhanced substantially at the onset of CDW transition, in good agreement with experimental observations in FeGe\cite{PCDai-Nature22}.
LC order generates orbital magnetic moments \cite{Park-PRB21, SZ-NC22, arXiv24-FuRQ} and may induce further change in magnetic moments.
It is desirable to conduct Kerr rotation, circular dichroism, muon relaxation, and nonlinear transport measurements to identify the nature of the CDW in FeGe.
We note, however, even if LC order were not realized in the real FeGe material, the interplay between magnetism and real bond-ordered CDW studied in this work continues to describe the essential observations of the experiments.

The physical origin of $2a_0 \times 2a_0$ CDWs in FeGe is still elusive. 
Recent experiments and theories \cite{YWang-PRM23, arXiv23-FengDL, arXiv23b-FengDL, JPLiu-CPL24, XGWan-CPL23, Lee-NC23, Chang-acs23, arXiv23-HXiang} suggest out-of-plane dimerization of Ge atoms that saves magnetic exchange energy, implying the importance of spin-phonon coupling, instead of electron-phonon coupling in \avs \cite{Ortiz-PRM19}.
On the other hand, neither electron-phonon nor spin-phonon coupling alone is capable of producing topological states with anomalous Hall effect.
The presence of vH singularities in the vicinity of Fermi level, together with the observed anomalous Hall effects in both FeGe and \avs, hint at possible connections between the CDWs in magnetic FeGe and nonmagnetic \avs\ kagom\'e metals.
The effective single-orbital $t$-$U$-$V_1$-$V_2$ model studied here unifies the physics of flat band magnetism, vH singularity CDW, and their interplay. 
It captures the essential physics of both \avs\ \cite{Dong-PRB23} and FeGe at suitable fillings and interactions, and has broad theoretical merits of its own that go beyond specific materials.

FeGe, as is \avs, is a multi-orbital material with multiple bands crossing the Fermi level, its more microscopic and quantitative descriptions would require further consideration of more realistic multi-orbital models.
We expect the main findings of this work to remain qualitatively valid, provided that the Fermi level lies inside the flat bands in the high-temperature symmetric phase and locates near the vH singularities in the low-temperature magnetically ordered state.
However, the leading CDW instability would depend on the detailed electronic structure.
For example, LC order is shown to be stabilized solely by nn $V_1$ in a multi-orbital model exhibiting multiple vH singularities close in energy and with opposite mirror eigenvalues \cite{Li-PRL24}. 

\textit{Acknowledgments.} -This work is supported by the National Key Research and Development Program of China (Grant No. 2022YFA1403800 and 2023YFA1407300) and the National Natural Science Foundation of China (Grant Nos. 12374153, 12447101, and 11974362).
Z.W. is supported by the U.S. Department of Energy, Basic Energy Sciences (Grant No. DE-FG02-99ER45747).
Z.W. is on sabbatical at the Kavli Institute for Theoretical Sciences, Chinese Academy of Sciences (KITS-CAS) and would like to thank KITS-CAS
for hospitality.
Numerical calculations in this work were performed on the HPC Cluster of ITP-CAS.

\end{document}